\documentclass[useAMS,usenatbib]{mn2e}

\usepackage{amsfonts,amsmath,amssymb,bm,epsfig,url}
\usepackage[dvips]{color}
\usepackage[normalem]{ulem}  
\newcommand{\df}[2]{ \frac{\partial {#1}}{\partial {#2}} }
\def\xme{m_e c^2}

\allowdisplaybreaks

\title[Maximum mass of magnetised white dwarfs]{On the maximum mass of
  magnetised white dwarfs}

\author[D. Chatterjee, A. F. Fantina, N. Chamel,
 J. Novak and M. Oertel]{
 D. Chatterjee,$^{1,2}$\thanks{E-mail: dchatterjee@lpccaen.in2p3.fr} 
 A. F. Fantina,$^{3,4}$\thanks{E-mail: anthea.fantina@ganil.fr}
 N. Chamel,$^3$\thanks{E-mail: nchamel@ulb.ac.be}
 J. Novak$^1$\thanks{E-mail: jerome.novak@obspm.fr}
 and M. Oertel$^1$\thanks{E-mail: micaela.oertel@obspm.fr}\\
 $^1$ LUTH, Observatoire de Paris, PSL Research University, CNRS,
 Universit\'e Paris Diderot, Sorbonne Paris Cit\'e,\\ 5 place Jules
 Janssen, 92195 Meudon, France\\
 $^2$  Laboratoire de Physique Corpusculaire, ENSICAEN, 6 Boulevard
 Mar\'echal Juin, F-14050 Caen C\'edex, France\\ 
 $^3$ Institut d'Astronomie et d'Astrophysique, CP-226,
 Universit\'e Libre de Bruxelles (ULB), 1050 Brussels, Belgium\\
 $^4$ Grand Acc\'el\'erateur National d'Ions Lourds (GANIL), CEA/DRF -
 CNRS/IN2P3, Bvd Henri Becquerel, 14076 Caen, France
}

\date{\today}

\pagerange{\pageref{firstpage}--\pageref{lastpage}} \pubyear{2016}

\begin{document}
\label{firstpage}

\maketitle

\begin{abstract}
  We develop a detailed and self-consistent numerical model for
  extremely-magnetised white dwarfs, which have been proposed as
  progenitors of overluminous Type Ia supernovae. This model can
  describe fully-consistent equilibria of magnetic stars in axial
  symmetry, with rotation, general-relativistic effects and realistic
  equations of state (including electron-ion interactions and taking
  into account Landau quantisation of electrons due to the magnetic
  field).  We study the influence of each of these ingredients onto
  the white dwarf structure and, in particular, on their maximum
  mass. We perform an extensive stability analysis of such objects,
  with their highest surface magnetic fields reaching $\sim 10^{13}$~G
  (at which point the star adopts a torus-like shape). We confirm
  previous speculations that although very massive strongly magnetised
  white dwarfs could potentially exist, the onset of electron captures
  and pycnonuclear reactions may severely limit their stability.
  Finally, the emission of gravitational waves by these objects is
  addressed, showing no possibility of detection by the currently
  planned space-based detector eLISA.
\end{abstract}


\section{Introduction}
\label{s:intro}

White dwarfs (WDs) are the stellar remnants of low and intermediate
mass stars, i.e. stars with masses $\lesssim$ 10 M$_\odot$ (M$_\odot$
being the mass of our Sun) \citep{shapiro1983}. The interest in WD
properties and in particular their mass-radius relation has been
renewed by the recent discovery of overluminous type Ia supernova
(SNIa) \citep{howell2006,scalzo,maoz14}. The progenitors of such events are
thought to be ``super-Chandrasekhar'' WDs with a mass $>$ 2 M$_\odot$
\citep[see, e.g.,][]{hillebrandt2013}. These SNIa may result from the
merger of two massive WDs or from the explosion of rapidly
differentially rotating WDs \citep[see,
  e.g.,][]{howell2006}. Alternatively, it has been argued that WDs
endowed with a strong magnetic field could be potential candidates for
the progenitors of these peculiar SNIa \citep{KunduBani, DasBaniPRD,
  das2012b, DasBaniPRL, das2013b, dasetal2013}. As a matter of fact, 
very massive strongly magnetised WDs were proposed much
earlier by \citet{shul76}; this work, however, does not seem to have
attracted much attention. The existence of magnetised supermassive WD is
of interest not only for astrophysics, but also for cosmology since 
SNIa have been used as "standard candles" to measure cosmological
distances assuming a unique astrophysical scenario for these
events. 

The mass-radius relation for non-magnetic WDs was established way back
in \citeyear{Chandrasekhar} by the pioneering work of
\citeauthor{Chandrasekhar}, assuming a simple degenerate electron Fermi gas
equation of state (EoS) at zero temperature. Subsequently, other
works attempted to construct more realistic models of WDs. Apart from
the magnetic field, different effects could alter the Chandrasekhar
limit for the maximum mass of a WD:
\begin{itemize}
\item Inclusion of finite temperature \citep{Marshak} and
  slow rotation \citep{Ostriker} did not result in significant deviations
  from the Chandrasekhar WD mass-radius relation. 
\item The electrostatic interaction between electrons and ions
  introduced by \citet{HamadaSalpeter} was found to lower the electron
  pressure resulting in a softer EoS and hence leads to slightly less massive
  configurations. 
\item The effect of general relativity on the structure of
  non-magnetic WDs was found to be non-negligible, reducing by a
    few per cent the WD maximum mass \citep[see,
  e.g.,][]{shapiro1983, Ibanez1983, Ibanez1984, Rotondo, ChinPhysB,
    Boshkayev1503a, BeraBhatt2015}.
\end{itemize}

The existence of magnetised WDs, with magnetic fields
$B \gtrsim 10^6$~G \citep{shapiro1983}, first predicted in the 1940s
by \citet{blackett1947}, was confirmed in 1970 by
\citeauthor{kemp1970} \citep[see, e.g.,][for a review]{jordan2009}.
At the time of this writing, hundreds of magnetised WDs have already
been discovered \citep{kepler2013}. While surface magnetic fields up
to about $10^9$~G can been inferred from Zeeman spectroscopy and
polarimetry, or cyclotron spectroscopy \citep[see, e.g.,][for a
review]{wf2000}, internal magnetic fields are not directly accessible
by observations. Therefore, very strong magnetic fields in the core of
WDs, as high as $B \sim 10^{13}$~G \citep{fujisawa2012}, cannot be
ruled out from a simple estimate of the energetics (i.e. scalar virial
theorem) or the conservation of magnetic flux of the progenitor
star. The question therefore arises of how such a strong magnetic
field affects the structure of WDs.

The study of the mass-radius relation of a magnetised WD has a long
history and it was recognised early on that the impact of the magnetic
field on both the stellar radius and mass could be large. However,
simplifying assumptions have been made for convenience. For instance,
the pioneering work by \citet{Ostriker} considered a vanishing
magnetic field at the surface of the star and neglected any magnetic
field effect on the EoS as well as general relativistic effects and
electrostatic interactions. The works of \citet{Adam},
\citet{DasBaniPRD}, and \citet{KunduBani} focused on the effect of the
magnetic field on the EoS, taking into account the Landau quantisation
of the electron gas, but were based on a Newtonian description of the
star's structure in spherical symmetry, i.e. neglecting the
deformation of the star by the magnetic field. A similar approach was
followed by \citet{SuhMathews}, where, however, the (spherical)
stellar structure was calculated from the general relativistic
Tolman-Oppenheimer-Volkoff equations. The deformations of the star
under the influence of a strong magnetic field have been recently
studied by \citet{BeraBhatt2014} in the Newtonian framework.
\citet{DasBaniGR} and \citet{BeraBhatt2015} have computed the
mass-radius relation of magnetic WDs for different magnetic field
geometries within a general relativistic framework using the publicly
available \textsc{XNS} code\footnote{
  http://www.arcetri.astro.it/science/ahead/XNS/index.html}. On the
other hand, Landau quantisation effects were neglected.

According to all these previous studies, the maximum possible mass of
strongly magnetised WDs lies substantially above the Chandrasekhar
limit, and more importantly above the mass inferred from observations
of overluminous SNIa, see
also~\citet{Cheoun2013,Herrera2013,Herrera2014,Federbush2015,Belyaev2015}.
Leaving aside the origin of such strong magnetic fields -- more than
three orders of magnitude higher than currently observed fields --,
the question as to whether such stars can possibly exist still needs
to be further examined. \citet{Coelho, ChamelFantinaDavis,
  ChamelPRD90, ChamelPRD92} discussed various microscopic and
macroscopic instabilities. In particular, \citet{ChamelFantinaDavis,
  ChamelPRD90} pointed out that the stability of very massive
magnetised WDs will not only be limited by general relativity, but
also by electron captures by nuclei and pycnonuclear reactions in
the stellar core. These microscopic processes, whose onset depends on
the magnetic field strength \citep{ChamelPRD92}, should thus be
carefully considered in modelling massive magnetised WDs.

In this study, we address these stability issues. To this end, we
shall construct consistent global general relativistic models of
strongly magnetised WDs, taking into account the
magnetic field effects on both the EoS and the WD equilibrium
structure equations. This work follows directly from our previous
study of the structure of strongly magnetised neutron stars
\citep{Chatterjee}. We shall account for the onset of electron capture
and pycnonuclear reactions, including the shifts on the threshold
densities and pressures due to electron-ion interactions and to the
strong magnetic field. Recently, during the refereeing process of our
paper, a very similar independent study by
\citet{Otoniel} appeared in the literature.

The paper is organised as follows. In Sec.~\ref{s:eos} we describe the
realistic dense matter EoS we considered, taking into account both the
effects of electron-ion interactions and the presence of the magnetic
field. In Sec.~\ref{s:structure}, we summarise the modified Maxwell
equations and the hydrodynamic equations in presence of a magnetic
field in general relativity and Newtonian theory. In
Sec.~\ref{s:mmax}, we describe the results concerning the influence of
an extremely strong magnetic field on the WD
structure. Sec.~\ref{s:rotwd} describes the effects of rotation, when
combined with the magnetic field; Sec.~\ref{s:resu_eos} gives in
details the influence of different microphysical ingredients on the WD
model; Sec.~\ref{s:instab} discusses physical instabilities that may
occur within our equilibrium models and Sec.~\ref{s:gw} makes a short
assessment on the detectability of gravitational waves emitted by
strongly magnetised WDs. Finally in Sec.~\ref{s:conc}, we give our
conclusions. Further detail on our model can be found in the
Appendix.

\section{Equation of state of strongly magnetised white
  dwarfs}\label{s:eos}
 
The model adopted here was initially developed for
the description of the outer crust of strongly magnetised neutron stars
\citep{lai91, chapav12}. The interior of magnetic WDs is assumed to
be composed of fully ionised atoms. Moreover, we assume that the
internal temperature has dropped below the crystallisation temperature
$T_m$ so that ions are arranged on a regular crystal lattice. The
crystallisation temperature is defined as \citep[see,
  e.g.,][]{haensel2007}
\begin{equation}\label{eq:Tm}
T_m=\frac{e^2}{a_e k_\text{B} \Gamma_m}Z^{5/3} \, ,
\end{equation}
where $e$ is the proton electric charge, $a_e=(3/(4\pi n_e))^{1/3}$ is
the electron-sphere radius, $n_e$ is the electron number density,
$k_\text{B}$ is the Boltzmann's constant, $\Gamma_m$ is the Coulomb
coupling parameter at melting. For simplicity, we shall consider
crystalline structures made of only one type of ions $^{A}_{Z}X$, with
mass number $A$ and atomic number $Z$ (typically carbon or oxygen). In
unmagnetised matter, $\Gamma_m\approx 175$ \citep{haensel2007}. In the 
presence of a strong magnetic field, the crystalline structure is more stable 
since $\Gamma_m\lesssim 175$ \citep{Potekhin2013}. 
Because the electron Fermi temperature is typically much smaller than $T_m$,
electrons are highly degenerate. In the following, we shall therefore
neglect thermal effects. The matter energy density is given by
\begin{equation}
\label{eq:eovera}
\mathcal{E} = n \frac{M^\prime(A,Z) c^2}{A} + \mathcal{E}_e +
\mathcal{E}_L - n_e m_e c^2 \ , 
\end{equation}
where $M^\prime(A,Z)$ is the mass of the nucleus $^A_ZX$ (including
the rest mass of nucleons and $Z$ electrons), $c$ is the speed of
light, $n$ is the baryon number density, $\mathcal{E}_e$ the energy
density of electrons, $m_e$ the electron mass, and $\mathcal{E}_L$ the
lattice energy density. The mass density $\rho$ is defined as
$\rho = m \, n$, where $m$ denotes the average mass per nucleon. The
last term in Eq.~(\ref{eq:eovera}), the electron rest mass energy, is
included to avoid double counting. For magnetic fields below
$\sim 10^{17}$~G, the nuclear masses remain essentially unchanged
\citep{pena2011,stein2016}. As in \citet{lai91} and \citet{chapav12}, we therefore
assume that nuclear masses are the same as in the absence of magnetic
fields. The nuclear mass $M^\prime(A,Z)$ can be obtained from the
tabulated atomic mass $M(A,Z)$ from the 2012 Atomic Mass Evaluation
\citep{audi2012} after subtracting out the binding energy of the
atomic electrons \citep[see Eq.~(A4) in][]{lpt03}:
\begin{eqnarray}
\label{eq:mprime}
M^\prime(A,Z) c^2 &=& M(A,Z)c^2 + 1.44381\times 10^{-5}\,Z^{2.39} \nonumber \\
   &+& 1.55468\times 10^{-12}\,Z^{5.35} \ ,
\end{eqnarray}
where both masses are expressed in units of MeV/c$^2$. 

In the presence of a strong magnetic field, the electron motion
perpendicular to the field is quantised into Landau levels \citep[see,
e.g., Chap.~4 in][]{haensel2007}. Ignoring the small electron
anomalous magnetic moment \citep[see, e.g., Section 4.1.1 in][and
references therein]{haensel2007}, and treating electrons as a
relativistic Fermi gas, the energies of Landau levels are given by
\begin{equation}
\label{eq:rabi}
\epsilon_{\nu} = \sqrt{c^2 p_z^2+m_e^2 c^4(1+2\nu_L b_{\star})}
\end{equation}
\begin{equation}
\nu_L = n_L + \frac{1}{2}+\sigma\, ,
\end{equation}
where $n_L$ is any non-negative integer, $\sigma=\pm 1/2$ is the spin,
$p_z$ is the component of the momentum along the magnetic field, and 
$b_{\star}=b/b_\textrm{crit}$ represents the magnetic field strength $b$ 
in units of the 
critical magnetic field $b_\textrm{crit}$ defined by
\begin{equation}
\label{eq:Bcrit}
 b_\textrm{crit}=\frac{m_e^2 c^3}{e\hbar}\approx 4.4\times 10^{13}~\text{G}\, .
\end{equation} 

For a given magnetic field strength $b_{\star}$, the number of occupied
Landau levels is determined by the electron number density $n_e$
\begin{equation}
\label{eq:ne}
n_e =\frac{2 b_{\star}}{(2 \pi)^2 \lambda_e^3} \sum_{\nu_L=0}^{\nu_L^{\rm
    max}} g_\nu x_e(\nu_L)\, ,
\end{equation}
\begin{equation}
\label{eq:xe}
x_e(\nu_L) =\sqrt{\gamma_e^2 -1-2 \nu_L b_{\star}}\, ,
\end{equation}
where $\lambda_e=\hbar/m_e c$ is the electron Compton wavelength,
$\gamma_e$ is the electron Fermi energy in units of the electron rest
mass energy,
\begin{equation}
\label{eq:gammae}
\gamma_e =\frac{\mu_e}{m_e c^2}\, ,
\end{equation}
with $\mu_e=d\mathcal{E}_e/dn_e$, while the degeneracy $g_\nu$ of a Landau 
level is $g_\nu=1$ for $\nu_L=0$ and $g_\nu=2$ for $\nu_L \geq 1$. 

The electron energy density $\mathcal{E}_e$ and corresponding electron
pressure $P_e$ are given by
\begin{equation}
\label{eq:Ee}
\mathcal{E}_e=\frac{b_{\star} m_e c^2}{(2 \pi)^2 \lambda_e^3}
\sum_{\nu_L=0}^{\nu_L^{\rm max}}g_\nu(1+2\nu_L b_{\star}) \psi_+
\biggl[\frac{x_e(\nu_L)}{\sqrt{1+2\nu_L b_{\star}}}\biggr]\, , 
\end{equation}
and 
\begin{equation}
\label{eq:Pe}
P_e=\frac{b_{\star} m_e c^2}{(2 \pi)^2 \lambda_e^3}
\sum_{\nu_L=0}^{\nu_L^{\rm max}}g_\nu(1+2\nu_L b_{\star}) \psi_-
\biggl[\frac{x_e(\nu_L)}{\sqrt{1+2\nu_L b_{\star}}}\biggr]\, , 
\end{equation}
respectively, where
\begin{equation}
\label{eq:psi}
\psi_\pm(x)=x\sqrt{1+x^2}\pm\ln(x+\sqrt{1+x^2})\, .
\end{equation}

A magnetic field will be referred to as strongly quantising if only the lowest level
$\nu_L=0$ is filled, or equivalently whenever $n_e<n_{e\text{b}}$, where 
\begin{equation}
\label{eq:neB}
n_{e\text{b}}= \frac{b_{\star}^{3/2}}{\sqrt{2} \pi^2 \lambda_e^3} \, ,
\end{equation}
which corresponds to the mass density 
\begin{equation}
\label{eq:rhoB}
\rho_{\rm b} = \frac{m}{y_e} \frac{b_{\star}^{3/2}}{\sqrt{2} \pi^2 \lambda_e^3} \, ,
\end{equation} 
where $y_e = n_e/n$.
Conversely, for a given mass density $\rho$ the strongly quantising
regime corresponds to magnetic field strengths 
\begin{equation}
b_{\star} >\left(\frac{\rho y_e \lambda_e^3
    \sqrt{2}\pi^2}{m}\right)^{2/3}\approx 180 (2 y_e \rho_{10})^{2/3},  
\end{equation} 
where $\rho_{10}=\rho/(10^{10}$~g~cm$^{-3}$). 

\begin{figure}
  \begin{center}
    \includegraphics[width=.4\textwidth]{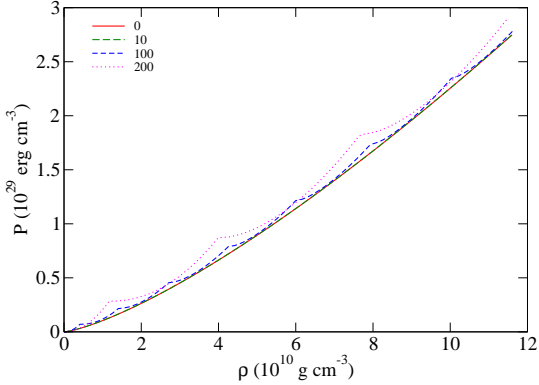}
    \caption{Matter contribution to the EoS (pressure $P$ vs mass density
      $\rho$) for a $^{12}$C WD, for
      different magnetic field strengths $b_{\star}$, where
      $b_{\star}=b/b_\textrm{crit}$ and $b_\textrm{crit}$ is defined
      in Eq.~(\ref{eq:Bcrit}).}
    \label{fig:wdeos1}
  \end{center}
\end{figure}

According to the Bohr-van Leeuwen theorem \citep{vanvleck1932}, the
lattice energy density is independent of the magnetic field, neglecting 
the small contribution due to the quantum zero-point motion
of ions \citep{baiko2009}. Considering point-like ions arranged in a
body-centred-cubic (bcc) structure~\citep{kozhberov2016}, the lattice 
energy density is given by 
\begin{equation}
\label{eq:EL}
\mathcal{E}_L  = \mathcal{C} e^2 n_e^{4/3} Z^{2/3} \ ,
\end{equation}
where $\mathcal{C} \approx -1.44$ \citep{baiko2001}. 
The corresponding lattice contribution to the pressure is given by
\begin{equation}
\label{eq:PL}
 P_L=\frac{\mathcal{E}_L}{3} \ .
\end{equation}
The matter contribution to the pressure is therefore (note that at zero
temperature nuclei do not contribute to the pressure): 
\begin{equation}
\label{eq:pmat}
P = P_e + P_L \ .
\end{equation}
The matter pressure $P$ is plotted in Fig.~\ref{fig:wdeos1} as a function 
of the mass density $\rho$ for different 
core magnetic field strengths $b_{\star}$ for a WD composed of $^{12}$C. 
The kinks correspond to the complete filling of Landau levels. 
From this figure it can already be anticipated that the
effect of the magnetic field on the EoS will have only a very small
influence on the WD structure, except for magnetic fields
$b_{\star} \gg 1$.

The magnetisation $\mathcal{M}$ is defined as:
\begin{equation}
\label{eq:M}
{\mathcal M} \equiv \left.\frac{\partial P}{\partial b}\right|_\mu \ .
\end{equation}
In the recent work of \citet{Otoniel}, nuclei are supposed to be
arranged in a simple cubic lattice. However, such a structure is known
to be unstable \citep{Born1940}. The baryon chemical potential $\mu$,
which coincides with Gibbs free energy per nucleon
\begin{equation}
\label{eq:gibbs}
g = \frac{\mathcal{E}+P}{n} \ ,
\end{equation}
is given by 
\begin{equation}
\label{eq:mu}
\mu = \frac{M^\prime(A,Z) c^2}{A} + \frac{Z}{A} \left( \mu_e -m_e c^2 +\mu_L \right) \ ,
\end{equation}
where 
\begin{equation}
\label{eq:mul}
\mu_L = \frac{4}{3} \frac{\mathcal{E}_L}{n_e} \ .
\end{equation}
Complete general expressions of the magnetisation as well as of its
derivatives can be found in Appendix~\ref{a:App-EoS}.

In the following, we shall consider stellar cores made of either 
$^{12}$C or $^{16}$O. These elements are the most likely fusion products 
of the helium burning phase.

\section{Stellar structure equations}\label{s:structure}

In this section, we outline the formalism to construct equilibrium WD
configurations starting from the EoS described in Sec.~\ref{s:eos}.
As general relativity has been shown to have a non-negligible effect
on the maximum WD mass, we compute the structure of WDs within both
Newtonian theory of gravity and the general relativistic framework,
for comparison (see also Sec.~\ref{s:gr}). The matter properties enter
the energy-momentum tensor which acts as a source of the Einstein
equations. The equilibrium is determined by the conservation of energy
and momentum, derived from the condition of vanishing divergence of
the energy-momentum tensor. We then compute equilibrium numerical
models of WDs to obtain the global structure properties such as mass
or radius.

The Einstein-Maxwell and equilibrium equations described hereafter are
solved using spectral methods within the numerical library
\textsc{lorene}\footnote{ http://www.lorene.obspm.fr}. We apply the
numerical scheme originally designed for strongly magnetised neutron
stars \citep[see][for further details]{Chatterjee} to study strongly
magnetised WD. As elaborated in \citet{Bocquet}, the electromagnetic
field tensor can be chosen to be either purely poloidal or purely
toroidal. Within the numerical scheme of \textsc{lorene}, the magnetic
field configuration is purely poloidal by construction. This is
convenient for comparison with the observed surface poloidal fields as
estimated from the spin-down measurements of pulsars. However, for
neutron stars, it has been suggested that amplification of the seed
poloidal magnetic field could result in the generation of strong
toroidal fields. Following the same arguments, one may argue that the
same holds for strongly magnetised WDs. Further, both purely poloidal
and toroidal magnetic field configurations are unstable, resulting in
the rearrangement of the configuration to a mixed one. That means, a
purely poloidal magnetic field configuration is not necessarily the
most general one. Studies of magnetised WDs in a general magnetic
field configuration have been recently performed employing the
\textsc{XNS} code \citep{BeraBhatt2014,BeraBhatt2015,DasBaniGR}. These
studies showed that large WD masses can be supported by a purely
toroidal field, but these configurations are also unstable. For the twisted 
torus mixed configuration (with a dominant poloidal
field), the mass-radius relations obtained were similar to those of
the purely poloidal case.
  
\subsection{Energy-momentum tensor in presence of a magnetic
  field}\label{s:emtensor}
In \citet{Chatterjee}, some of us explicitly derived the
generalised expression for the thermodynamic average of the
microscopic energy-momentum tensor, which is required for the study of
the macroscopic structure of a compact object, as recapitulated below:
\begin{eqnarray}
\langle T^{\mu\nu}\rangle  &=& (\mathcal{E} + P)\; u^\mu u^\nu + P
  \; g^{\mu\nu} \nonumber \\ && + \frac{1}{2} (F^\nu_{\;\tau} {\mathcal M}^{\tau\mu} +
F^{\mu}_{\;\tau} {\mathcal M}^{\tau\nu} ) \nonumber \\ && 
 - \frac{1}{\mu_0} (F^{\mu\alpha} F_{\alpha}^{\,\nu} + \frac{g^{\mu\nu}}
{4} F_{\alpha\beta} F^{\alpha\beta}) ~.
\label{eq:tmunu}
\end{eqnarray}
where, $\mathcal{E}$ is the matter energy density, $P$ is the matter pressure
(see Sec. \ref{s:eos}), $u^{\mu}$ is the fluid four-velocity and
$\mu_0$ is the vacuum magnetic permeability in S.I. units. The
electromagnetic field strength tensor is derived from the
electromagnetic potential 1-form $A_\mu$ through
\begin{equation}\label{e:def_Fmunu}
F_{\mu\nu} = \frac{\partial A_\nu}{\partial x^\mu} - \frac{\partial
  A_\mu}{\partial x^\nu}\, ,
\end{equation}
and ${\mathcal M}^{\mu \nu}$ is the magnetisation tensor (see
\citet{Chatterjee} for complete derivation). 

The first two terms on the right hand side of \eqref{eq:tmunu} can be
identified as the pure (perfect fluid) matter contribution, followed
by the magnetisation term and finally the usual electromagnetic field
contributions to the energy-momentum tensor. For isotropic media, the
magnetisation is aligned with the magnetic field. Thus one may write
$\displaystyle {\mathcal M}_{\alpha \beta} = \frac{\chi}{\mu_0}
F_{\alpha \beta}$, in terms of the dimensionless scalar 
\begin{equation}
\chi = \frac{\mu_0 {\mathcal M}}{b}\, .\label{e:def_chi}
\end{equation}
Finally, the electric and magnetic fields as measured by the Eulerian
observer have only two non-vanishing components each \citep[see][]{BGSM}:
\begin{subequations}\label{e:def_EB}
  \begin{eqnarray}
  E_r & = &
  \frac{1}{N}\left( \df{A_t}{r} + N^\varphi \df{A_\varphi}{r}
  \right), \label{e:def_Er}\\
  E_\theta & = & \frac{1}{N}\left( \df{A_t}{\theta} + N^\varphi \df{A_\varphi}{\theta}
  \right), \label{e:def_Et}\\
  B_r & = &
  \frac{1}{C r^2\sin\theta}\df{A_\varphi}{\theta}, \label{e:def_Br}\\
  B_\theta & = & - \frac{1}{C \sin \theta} \df{A_\varphi}{r}, \label{e:def_Bt}
\end{eqnarray}
\end{subequations}
where $C, N$ and $N^\varphi$ are gravitational potentials defined by
the metric~\eqref{e:metric}. Note that we denote the magnetic field
norm in the matter comoving frame by $b$, to distinguish it from that
in the Eulerian frame, denoted by $B$ with components $B_r$ and
$B_\theta$.

\subsection{Einstein-Maxwell and equilibrium
  equations}\label{s:einsteinmaxwell}

To construct numerical models of magnetic WDs, we adapt the general
relativitic scheme following previous works by
\citet{Bocquet,BGSM}. Under the assumptions of stationarity,
axisymmetry and circularity of the spacetime, we employ the maximally
sliced quasi-isotropic coordinates in which the line element can be
written as:
\begin{eqnarray}
  {\rm d}s^2 &=& -N^2\, {\rm d}t^2 + C^2r^2\sin^2\theta \left({\rm d}\varphi -
    N^\varphi\, {\rm d}t \right)^2 \nonumber\\
  &&+ D^2\left( {\rm d}r^2 + r^2\, {\rm d}\theta^2\right),\label{e:metric}
\end{eqnarray}
where $N, N^\varphi, C$ and $D$ are functions of coordinates
$(r, \theta)$. With these properties and coordinate choice, the
Einstein equations reduce to a set of four elliptic partial
differential equations for the four gravitational potentials, in
which part of the source terms is derived from the energy-momentum
tensor. These equations are then solved for a given matter
content. Note that, in this section, Latin letters i, j,~\dots are used
for spatial indices only, whereas Greek ones $\alpha$, $\mu$,~\dots
denote the spacetime indices.

Whereas the homogeneous Maxwell equation is automatically verified by
the expression~\eqref{e:def_Fmunu} for the electromagnetic field
tensor, the inhomogeneous one must be modified to include the
contribution from the magnetisation \citep{Chatterjee}:
\begin{equation}
\frac{1}{\mu_0} \nabla_\mu F^{\nu\mu} = j_{\ \mathit{free}}^\nu +
\nabla_\mu {\mathcal M}^{\nu\mu}~, \label{e:Maxwell}
\end{equation}
where $j_{\ \mathit{free}}^\nu$ is the free current which generates
the electromagnetic field, as opposed to the bound current
contribution associated with the magnetisation (last term in the above
equation). For a discussion about these free currents, which are not
driven by the bulk motion of matter, see \citet{BGSM}.

To obtain equilibrium configurations, one must solve the
coupled Einstein-Maxwell equations, together with the condition of
conservation of energy and momentum $\nabla_\mu T^{\mu\nu} = 0$,
giving rise to the equilibrium condition in our stationary case. It
was already illustrated in \citet{Chatterjee} that on inclusion of the
magnetisation current density in the equation of 
equilibrium, the Lorentz force associated with the magnetisation
current cancels with the magnetisation contribution arising from the
pressure gradient (key equations are recapitulated in the Appendix
\ref{a:App}). Therefore, there is no explicit magnetisation appearing
in the equilibrium condition, and the first integral of
fluid stationary motion remains unchanged retaining the same form as
in the case without magnetisation (see details in App.~\ref{a:App}):
\begin{equation}
  \label{eq:1st_integral}
  \ln h(r, \theta) + \nu(r, \theta) - \ln \Gamma(r, \theta) +
  \Phi(r,\theta) = {\rm const}, 
\end{equation}
where $h$ is the enthalpy (as we assume zero temperature, $h = g$ the
Gibbs free energy, introduced in Eq.~\eqref{eq:gibbs}), $\nu = \log N$
a gravitational potential function, $\Gamma$ is the Lorentz factor and
$\Phi$ is the electromagnetic term associated with the Lorentz force
\citep[see][for a discussion]{BGSM}:
\begin{equation}
  \label{e:def_Phi}
  \frac{\partial \Phi}{\partial x^i} = - \frac{F_{i\mu} j^\mu_{\
      \mathit{free}}}{\mathcal{E} + P}.
\end{equation}

\subsection{Newtonian limit}\label{s:nt}
As WDs are not strongly relativistic objects, their structure
is often calculated in a non-relativistic Newtonian framework. We
summarise here the basic stellar structure equations that the general
relativistic equations reduce to in the Newtonian limit.

Whereas for Maxwell equations one must consider the
special-relativistic form, the Newtonian limit of the Einstein
equations is the Poisson equation for the gravitational potential
$\phi$:
\begin{equation}
\nabla^2 \phi = 4 \pi G \rho~,
\end{equation}
where $G$ is the gravitational constant and $\rho$ is the mass density.
Thus in the non-relativistic limit, the factor $\nu$ tends to the
Newtonian gravitational potential $\phi$ \citep{GourgoulhonLect}. 

The first integral of motion (Eq. \ref{eq:1st_integral}), described in
the previous section, reduces to a simple form in the Newtonian limit
in the absence of an electromagnetic field
\begin{equation} 
\label{eq:1st_integral_newt}
\hat{h} + \phi - \frac{1}{2} U^2 = {\rm const},
\end{equation}
where $\hat{h} = \left( \hat{\mathcal{E}} + P \right) / \rho$ is the
non-relativistic enthalpy (\textit{i.e.} $\mathcal{E} = \rho\,c^2 +
\hat{\mathcal{E}}$, $\hat{\mathcal{E}}$ being the internal energy, not
taking into account rest-mass energy density) and $U$ the fluid
velocity in the $\varphi$-direction. Note here that, as shown,
e.g., by \citet{GourgoulhonLect} this is not the classical
Bernoulli equation, but a first integral of motion, which is valid
only if the fluid is in pure circular motion, while the Bernoulli
theorem is valid for any stationary flow. Moreover, the expression for
the Bernoulli theorem provides a constant along each fluid line, which
may vary from one fluid line to another, whereas the first
integral~\eqref{eq:1st_integral_newt} gives a constant valid in the
entire star.

In presence of free currents $j^{\mu}_{\ \mathit{free}}$, see
Eq.~\eqref{e:Maxwell} and of the electromagnetic field they generate,
a force term of the form
$\displaystyle f_i = F_{i \sigma} j^{\sigma}_{\ \mathit{free}}$ is
added to the equation of stationary motion \citep{BGSM}. Now, using
the definition of $F_{\mu\nu}$~(\ref{e:def_Fmunu}), together with the
Newtonian limit (\textit{i.e.} the metric potentials $N, C, D\to 1$
and $N^\varphi \to 0$) in Eqs.~\eqref{e:def_EB}, and writing the
4-currents $j^\mu$ as a charge density $j^0_{\ \mathit{free}}$ and a
current vector $\vec{j}_{\mathit{free}}$ separately, one gets the
standard form for the force:
\begin{equation}
  \label{e:EM_force}
  \vec{f} = j^0_{\ \mathit{free}}\, \vec{E}  + \vec{\jmath}_{\mathit{free}} \times \vec{B}.
\end{equation}
As we assume the matter inside the star to be a perfect conductor, we
have the relation:
\begin{equation}
  \label{e:perfect_conductor_Newt}
  \vec{E} = - \vec{v} \times \vec{B},
\end{equation}
with $\vec{v} = \Omega\, \vec{e}^\varphi$ the fluid velocity
\footnote{$\Omega$ is the angular velocity and $\vec{e}^\varphi$ is
  the local triad vector associated with the $\varphi$-coordinate.}
(thus, $U= \Omega r\sin \theta$) . The Newtonian interpretation of our
requirement of having a first integral of motion, with the
introduction of the potential $\Phi$~\eqref{e:def_Phi}, is that the
electromagnetic force be potential-like:
\begin{equation}
\label{e:1st_integ_Newt}
\vec{f} = \left( \vec{\jmath}  - \Omega\, \vec{e}^\varphi j^0 \right)
\times \vec{B} = \rho \vec{\nabla} \Phi.
\end{equation}
The equilibrium equation in the Newtonian case then reads:
\begin{equation}
  \label{e:Newt_MHD}
  \hat{h} + \phi - \frac{1}{2} U^2 + \Phi = \textrm{const}.
\end{equation}
As demonstrated before in the relativistic case, note again that the
above equations for equilibrium do not contain any
contribution from the magnetisation. They thus differ from those given
in \citet{BeraBhatt2014}, where the magnetisation has been artificially
included.

\section{Stellar structure and magnetic field}
\label{s:mmax}

With the numerical setup described in Sec.~\ref{s:structure}, we
construct models of magnetic $^{12}$C WDs in equilibrium and determine
their structure for various magnetic field strengths. After comparing
our results with those obtained in previous studies, we show that the
the relevant quantity to consider for assessing the global stability
of strongly magnetised WDs is the magnetic dipole moment rather than
the magnetic field strength. We compute the maximum mass and discuss
the limitations of our model. Since the magnetic field dependence of
the EoS is very weak, see the discussion in Sec.~\ref{sec:eosmag}, we
will neglect unless otherwise stated magnetisation and magnetic field
dependence of the EoS for simplicity within the calculations.

\subsection{Role of a strong magnetic field on the white dwarf structure}
\label{ss:B_effect}

Based on a non-rotating model in Newtonian gravity, we first vary
central enthalpy along sequences of fixed central magnetic fields
$B_{\star}$ where $B_{\star}=B/b_\textrm{crit}$, $b_\textrm{crit}$
being defined in Eq.~(\ref{eq:Bcrit}) and $B$ is the magnetic field
measured by the Eulerian observer. The value remaining fixed along a
given sequence corresponds to the value of magnetic field at the
centre of the star. Indeed, for each choice of free current
distribution~(\ref{e:Maxwell}) one obtains a given magnetic field
distribution inside the star and thus one can determine the value of
central magnetic field strength.
\begin{figure}
  \begin{center}
      \includegraphics[width=.4\textwidth,angle=270]{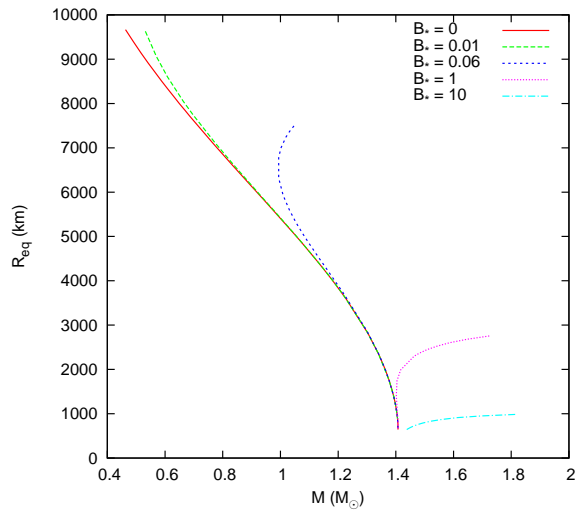}
      \caption{Masses $M$ in units of solar mass (M$_\odot$) as
        functions of equatorial radii $R_{eq}$ for $^{12}$C WDs in
        Newtonian gravity along 5 sequences of fixed central magnetic
        field $B_{\star}$, ranging from $0$ to
        $10$. $B_{\star}=B/b_\textrm{crit}$ and $b_\textrm{crit}$ is
        defined in Eq.~(\ref{eq:Bcrit}).}
    \label{fig:mr_constb}
  \end{center}
\end{figure}

\begin{figure}
  \begin{center}
       \includegraphics[width=.4\textwidth,angle=270]{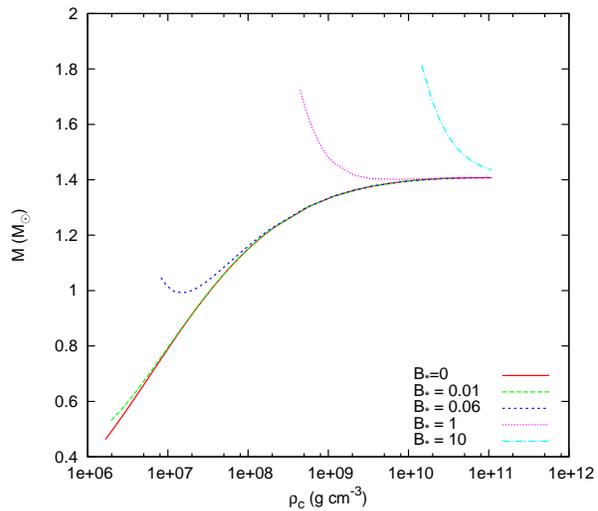}
       \caption{Masses $M$ in units of solar mass (M$_\odot$) as
         functions of central density $\rho_c$ for $^{12}$C WDs in
         Newtonian gravity along 5 sequences of fixed central magnetic
         fields $B_{\star}$, ranging from $0$ to
         $10$. $B_{\star}=B/b_\textrm{crit}$ and $b_\textrm{crit}$ is
         defined in Eq.~(\ref{eq:Bcrit}).}
    \label{fig:rhomg_constb}
  \end{center}
\end{figure}

Fig.~\ref{fig:mr_constb} displays the mass $M$ (in units of solar mass
M$_\odot$) as a function of equatorial radius $R_{eq}$ (in km) for such
sequences in Newtonian gravity. For low magnetic fields, the sequences
follow the mass-radius relation as for non-magnetic WDs, ultimately
reaching the Chandrasekhar limit. As the field strength is increased,
the deviation from the non-magnetic curve increases, resulting in
configurations with higher mass at lower densities. This is clearly
evident from Fig.~\ref{fig:rhomg_constb}, where for the same sequences
we show the masses for varying central mass density $\rho_c$ (in g
cm$^{-3}$). These results in Newtonian gravity are well known and can
be compared with those of \citet{SuhMathews,BeraBhatt2014} and are
found to be in accordance. For field strengths of about 10
$b_{\rm crit}$, masses as large as $\sim$ 1.8~M$_{\odot}$ are found to be
supported by the magnetic field for both $^{12}$C and $^{16}$O.

These figures clearly demonstrate that WD masses larger than the
standard Chandrasekhar limit for the non-magnetic case can be
supported by strong magnetic fields. However, the strongly magnetised
sequences ($B_\star = 1, 10$) seem to be gravitationally unstable, if
one uses the standard criterion for non-magnetic stars
$\partial R_{eq} / \partial M > 0$ or
$\partial M / \partial \rho_c < 0$ \citep[e.g.][]{shapiro1983}. The
important point here is that, in order to look for such instabilities
along families of two-parameter equilibria, one must be careful in
choosing the quantity that should be kept constant when varying the
mass \citep{sorkin82}. Indeed, just as in the case of a rotating body
it is the angular momentum and not the angular velocity that is the
constant of motion, it can be shown that the conservation of magnetic
flux implies that magnetic moment is a constant of motion and not the
magnetic field. Therefore, sequences of fixed dipole magnetic moments
$\mathcal{D}$ should be considered, and not of fixed central magnetic
field value as in Figs.~\ref{fig:mr_constb} and
\ref{fig:rhomg_constb}. This dipole moment is computed from the
asymptotic behaviour of the magnetic field, obtained from our
numerical model \citep[see][for a definition]{Bocquet}.

\begin{figure}
  \begin{center}
      \includegraphics[width=.4\textwidth,angle=270]{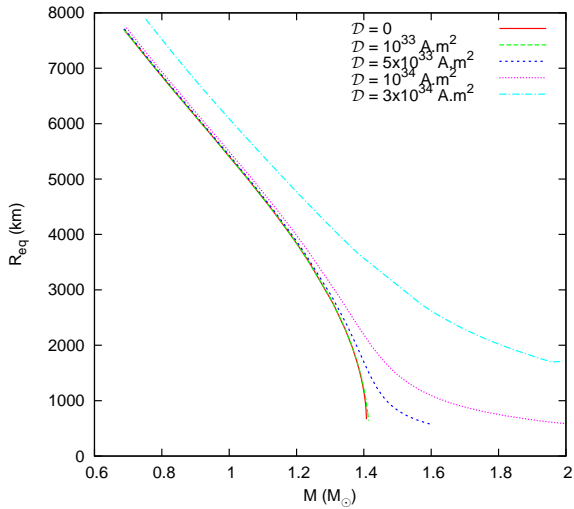}
      \caption{$^{12}$C WD masses $M$ vs radii $R_{eq}$
        along sequences of fixed magnetic moments $\mathcal{D}$ in
        Newtonian gravity.}
    \label{fig:mr_constmm}
  \end{center}
\end{figure}

In Fig.~\ref{fig:mr_constmm} we therefore construct mass-radius
relations for increasing central enthalpy along sequences of fixed
dipole magnetic moments $\mathcal{D}$. As seen from this figure, even
very highly-magnetised sequences, which can reach two solar masses
configurations, are gravitationally stable in Newtonian
gravity. Effects of relativistic gravity (General Relativity) shall be
discussed later, in Sec.~\ref{s:gr}.

\begin{figure}
  \begin{center}
      \includegraphics[width=.4\textwidth,angle=270]{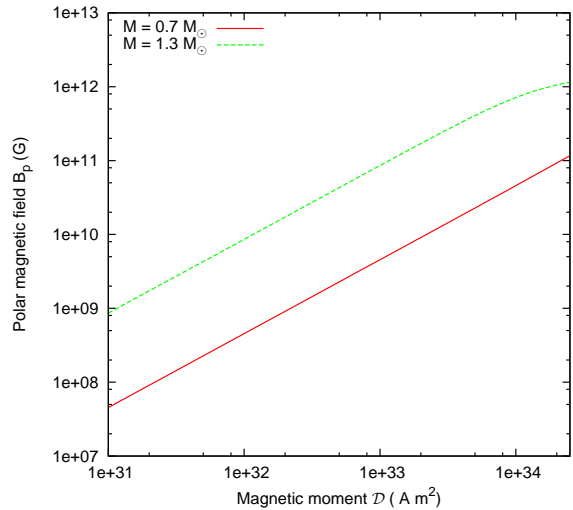}
      \caption{Polar magnetic field $B_P$ vs magnetic dipole moment
        $\mathcal{D}$ for $^{12}$C Newtonian WDs of baryon masses 0.7
        $M_{\odot}$ and 1.3 $M_{\odot}$}
    \label{fig:bpmm}
  \end{center}
\end{figure}

Finally, as it is not the magnetic moment which is an astrophysically
observable quantity, but rather the polar magnetic field which is
derived from the measurement of the spin and spin periods, we here
give relations between both quantities, so that one can get an idea of
typical magnetic fields arising for the magnetic dipolar moments
quoted in this study. For two exemplary typical values of WD
masses (0.7 and 1.3 solar masses), we display in Fig.~\ref{fig:bpmm}
the values of polar magnetic fields $B_P$ (in G) for a $^{12}$C WD
corresponding to the range of magnetic moments considered in this
work. One must here keep in mind that surface magnetic fields of
isolated magnetic WDs are observed to lie within the range $10^3$ -
$10^9$~G \citep{Ferrario}.

\subsection{Maximally distorted stellar configurations}\label{s:distort}

\begin{figure}
  \begin{center}
    \includegraphics[angle=-90,scale=.45]{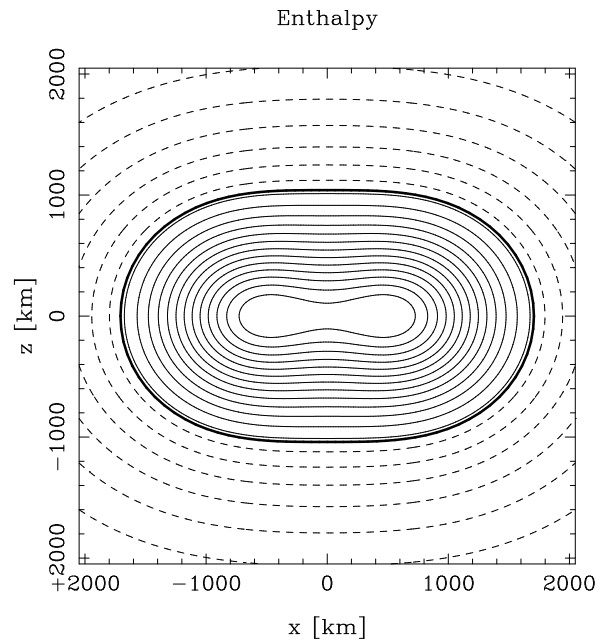}
  \end{center}
  \caption{Enthalpy isocontours of a static stellar configuration in
    the $(x,z)$ plane of a $^{12}$C WD for a magnetic dipole moment of
    $3 \times 10^{34}$ A m$^2$. The mass of the star is $1.99 M_\odot$
    and the polar magnetic field $B_P \sim 3 \times 10^{13}$~G.}
  \label{fig:fig7_mm3e34NT}
\end{figure}

With stronger magnetic fields, the anisotropic Lorentz force acting on
the free currents~(\ref{e:EM_force}) causes the stellar structure to
increasingly deviate from spherical symmetry, finally reaching a
toroidal shape. At this point the code reaches the limit of its
numerical capability for describing the stellar surface, and hence
more massive stable WD configurations cannot be achieved within this
numerical framework. As an example, in Fig.~\ref{fig:fig7_mm3e34NT} we
illustrate the maximally distorted shape of the stellar surface of a
$^{12}$C WD for a magnetic dipole moment of $3 \times 10^{34}$ A
m$^2$. The induced magnetic field at the pole rises up to
$3.1\times 10^{13}$~G (equivalently $B_\star \sim 1$) whereas the
magnetic field at the centre reaches $B_\star \sim 10$. The ratio of
magnetic to fluid pressure computed at the centre of the WD in this
case is found to be $\sim 0.79$, and $0.83$ in the $^{16}$O EoS
case. The endpoints of the curves for two-solar masses magnetised WDs
in Fig.~\ref{fig:mr_constmm} correspond to these maximally distorted
configurations, for which the numerical approach start being
inaccurate. Strictly speaking, this does not represent an instability,
and more massive stars with torus-like shape are not \textit{a priori}
excluded \citep[see also the discussion in the case of neutron stars
  by][]{CardallLattimerPrakash}. Nevertheless, it should be kept in mind
that for these highly magnetised stars, the polar field is four orders
of magnitude above the highest currently observed values and that, in
particular, it is not clear in which way such a huge magnetic field
could be formed within a WD. The stability of such toroidal stars has
not been checked, either.

Another point is that we considered here a purely poloidal magnetic
field configuration. Purely toroidal configurations might lead to much
higher masses. However, it remains to be shown that these configurations are stable. 
In addition, under the fossil field hypothesis, the
WD magnetic field is probably dominated by the poloidal field. As shown,
e.g. in \citet{BeraBhatt2015}, the results for poloidal dominated
mixed configurations are fairly similar to the results for purely
poloidal fields discussed here. In Sec.~\ref{s:instab} we will
discuss other instabilities that may limit the mass of a magnetised
WD.

\section{Stellar structure and rotation}\label{s:rotwd}

\begin{figure}
  \begin{center}
      \includegraphics[width=.4\textwidth,angle=270]{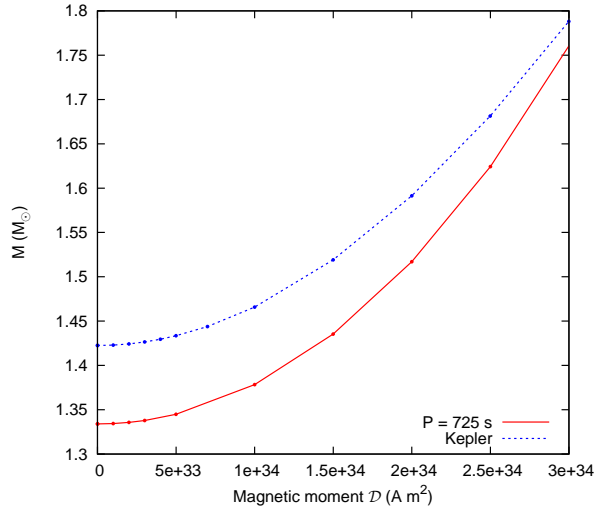}
      \caption{Masses for Newtonian magnetic WDs composed of $^{16}$O rotating with a period of 725~s (solid line)
        and at Kepler frequency (dashed line)
        as a function of fixed magnetic moments ${\mathcal{D}}$. 
        }
    \label{fig:mmmg_hc0.002_kep}
  \end{center}
\end{figure}

It is already a well known result that rotation provides centrifugal
force that can support larger masses in compact stars. We estimated
the relative increase in mass of a rotating magnetic WD compared to a
non-rotating one. The observed rotation periods of isolated magnetic
WDs span the range from 725~s to decades or centuries
\citep{Ferrario}. Magnetic cataclysmic variables exhibit shorter
periods, with AE Aqr having a period of 33~s. For such values of the
rotation period, we find that the relative difference between the
masses of non-rotating and rotating WDs at a given density and
magnetic dipole moment is negligible (of the order of $10^{-4}$ or
lower). On the contrary, the magnetic field plays a major role in
supporting massive WDs. To demonstrate this fact, we display in
Fig. \ref{fig:mmmg_hc0.002_kep} the maximum mass of a WD rotating with
a period of 33~s (depicted by a solid line) - these results are
indistinguishable from those obtained for a nonrotating WD - as well
as a WD rotating at the Kepler frequency corresponding to the mass
shedding limit (dashed line) for increasing values of the magnetic
dipole moment. It is evident that even for the fastest rotating
observed WDs, rigid rotation does not lead to any significant increase
in their mass. More importantly, the effects of rotation are
comparatively less important for strongly magnetised WDs, even at
Kepler frequency, as the main force resisting gravity is no longer the
centrifugal buoyancy, but the Lorentz force due to the magnetic field.

On the other hand, differentially rotating compact stars can be
significantly more massive than their non-rotating or uniformly
rotating counterparts \citep[see, e.g.,][]{Subra}. However, the support
due to differential rotation would ultimately be cancelled by magnetic
braking and/or viscosity. These processes drive the star into uniform
rotation \citep{Shapiro2000}. As a matter of fact, \citet{BGSM} showed
that a magnetic star in a stationary state is necessarily rigidly
rotating.

\section{Stellar structure and the equation of state}\label{s:resu_eos}

In this section, we perform a systematic analysis of the role of the EoS on the 
WD structure, and more specifically the effects of the magnetic field and 
electron-ion interactions. 

\subsection{Magnetic field dependence of the equation of state}
\label{sec:eosmag}
\begin{figure}
  \begin{center}
       \includegraphics[width=.4\textwidth,angle=270]{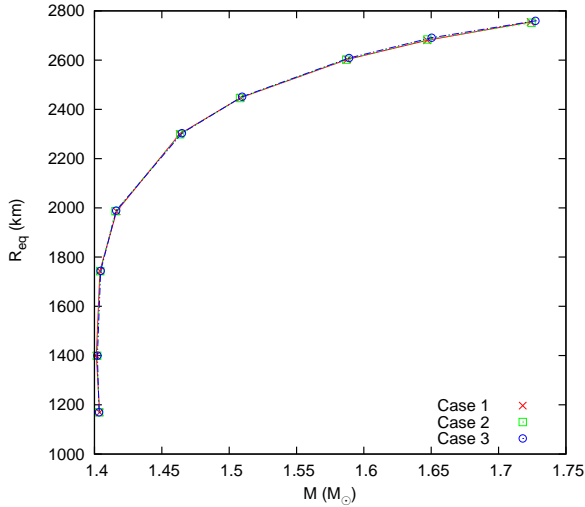}
       \caption{Radius $R_{eq}$ vs Mass $M$ for $^{12}$C
         Newtonian WDs for fixed central magnetic field
         $B_c = b_{\rm crit}$~(\ref{eq:Bcrit}), with and without
         magnetic field dependence in EoS and magnetisation. See text
         for details about the cases 1, 2 and 3.}
    \label{fig:mr_allcases}
  \end{center}
\end{figure}

As discussed in Section~\ref{s:eos}, the magnetic field influences the properties of dense 
matter in magnetic WDs through (i) Landau quantisation (i.e. magnetic-field dependent EoS), 
and (ii) the magnetisation $\mathcal{M}$. In order to assess the relative 
importance of these effects, we compute the structure of magnetic WDs 
for a fixed central magnetic field $B_c = b_{\rm crit}$ considering the 
three different cases: 
\begin{itemize}
\item Case 1: EoS ignoring magnetic field effects,
\item Case 2: EoS including Landau quantisation but neglecting magnetisation,
\item Case 3: EoS including both Landau quantisation and magnetisation.
\end{itemize}
As shown Fig.~\ref{fig:mr_allcases}, the corresponding mass-radius
relationships are practically indistinguishable, in accordance with
the results previously obtained by \citet{BeraBhatt2014}. Considering
this result, Fig.~\ref{fig:wdeos1}, the discussion in
Sec.~\ref{s:distort} and the fact that the maximum values of the
magnetic field inside the star remain low, only of the order of
$B_{\star} \lesssim 10$, even for the maximally distorted
configurations, we will adopt the simplest case 1 for the following
calculations.

\subsection{Electron-ion interactions}

\begin{figure}
  \begin{center}
      \includegraphics[width=.4\textwidth,angle=270]{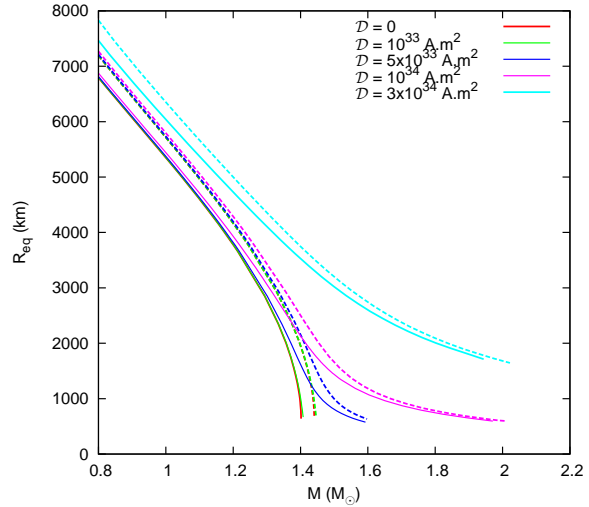}
      \caption{Mass-radius relations for
        Newtonian $^{16}$O WDs for different values of fixed
        magnetic moments with (solid lines) and without (dashed lines)
        lattice effects. 
}
    \label{fig:latcomp+pycno}
  \end{center}
\end{figure}

Most previous investigations of magnetic WDs have employed the EoS of
an electron Fermi gas
\citep{DasBaniPRD,KunduBani,BeraBhatt2014,DasBaniGR,Franzon2015}.
\citet{BeraBhatt2015} and \citet{Otoniel} have recently computed the
structure of magnetic WDs including the lattice correction to the EoS.
\citet{BeraBhatt2015} employed the expression obtained by
\citet{Salpeter} in the Wigner-Seitz approximation, while
\citet{Otoniel} used the expression for a simple-cubic
lattice. However, as already pointed out in Section~\ref{s:eos}, this
lattice type is unstable. In all our calculations presented so far,
(body-centred cubic) lattice corrections were taken into account. But
in order to assess the importance of these effects, we now set
$\mathcal{C}=0$, see Fig.~\ref{fig:latcomp+pycno}, where the results
without lattice effects are compared with the results taking them into
account. We find that electron-ion interactions lead to a lower
maximum mass for WDs, as previously discussed by \citet{ChamelPRD92}
for non-magnetic WDs. However, the reduction is rather small: the
maximum mass of $^{12}$C and $^{16}$O non-magnetic WDs thus decreases
from $\sim$ 1.44 M$_{\odot}$ to $\sim$ 1.41 M$_{\odot}$. On the other
hand, lattice effects can have a much larger impact on the stellar
radius, up to about 50\% for a non-magnetic star with a mass of 1.4
M$_{\odot}$. These effects are less pronounced in strongly magnetised
WDs.

\section{Stability of strongly magnetised white dwarfs}\label{s:instab}
In this Section, we discuss several physical mechanisms that are able
to affect the WD stability, turning some of the equilibrium solutions
we have found so far into unstable ones. The aspects we consider here
are the use of general relativity for the description of gravity;
electron-capture instability and pycnonuclear instability. A summary
of the different maximum mass values, considering the mechanisms
discussed in this Section, can be found in Table~\ref{tab:mmax_O16}
for $^{16}$O WDs and in Table~\ref{tab:mmax_C12} for $^{12}$C WDs.
  
\subsection{General-relativistic instability}\label{s:gr}

\begin{figure}
  \begin{center}
      \includegraphics[width=.4\textwidth,angle=270]{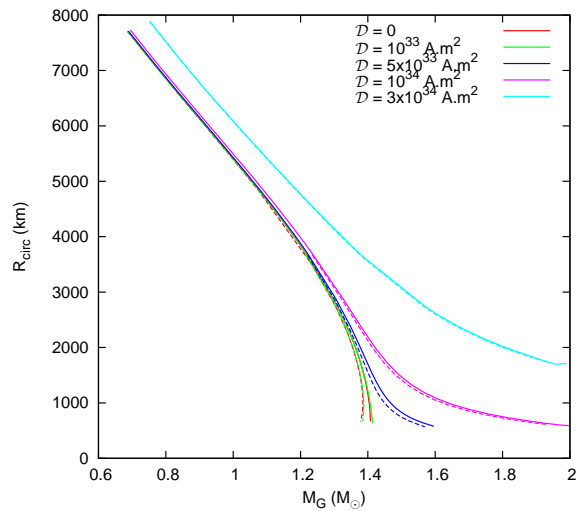}
      \caption{Circumferential radius $R_{circ}$ vs gravitational mass
        $M_G$ for magnetic $^{12}$C WDs along sequences of fixed
        magnetic dipole moments $\mathcal{D}$ in Newtonian theory
        (bold lines) and general relativity (dashed lines).}
    \label{fig:mr_constmm_compGR}
  \end{center}
\end{figure}

Although most mass-radius models for WDs have been constructed in
Newtonian gravity, it is well known that computing equilibrium
configurations within general relativity reduces the maximum mass for
the case of non-magnetic
WDs~(e.g. \citet{Ibanez1983,Ibanez1984,Rotondo,Boshkayev1503a}). We
compute the structure of magnetised WDs both in Newtonian theory and
in general relativity. To estimate how general relativistic effects
may limit the maximum masses, we plot in
Fig.~\ref{fig:mr_constmm_compGR} the mass-radius relationship for the
non-magnetic as well as the magnetic WDs considering both Newtonian
theory and general relativity. The mass used in general relativity is
the so-called \textsl{gravitational mass}, which is the mass felt by a
test-particle orbiting around the WD. For a complete definition,
details of its computation and the definition of the circumferential
equatorial radius used there too, see~\citet{BGSM}. It is clear from
this figure that general relativity has a non-negligible effect in
limiting the maximum mass of non-magnetic WD. We observed that on
inclusion of general relativistic effects, the maximum mass of WD was
reduced from the 1.41 M$_{\odot}$ to 1.38 M$_{\odot}$.

For the case of strongly magnetised WDs, it is found that it is not
general relativity that plays a crucial role in limiting the maximum
masses but the magnetic field which distorts the WD beyond the
equilibrium poloidal shape. The reason is that general relativity
affects the high-density part of the EoS, which determines the maximum
mass for the non-magnetic case, but for the magnetic case it is rather
the low-density part of the EoS which determines the maximum mass. To
understand this we recall Fig.~\ref{fig:rhomg_constb}, where we
plotted the masses as a function of density, from which it was evident
that it is the low density part of the curves which bends away from
the non-magnetic mass-radius relation and are responsible for the
large masses, and the high density part remains unaffected. That the
effect of general relativity on the maximum masses is small ($ < 2\%$
for poloidal fields) for the case of strongly magnetised WDs was
demonstrated earlier by \citet{DasBaniGR,BeraBhatt2015} employing the
\textsc{XNS} and \textsc{LORENE} codes.

\subsection{Electron capture instability}\label{s:ec}

\begin{figure}
  \begin{center}
      \includegraphics[width=.4\textwidth,angle=270]{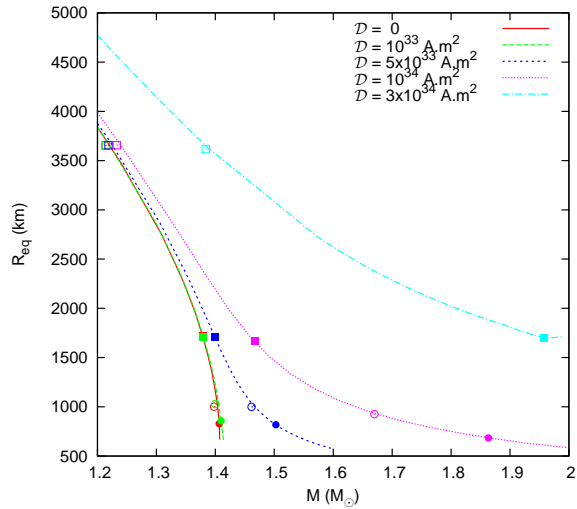}
      \caption{Mass radius relationship for magnetic WDs in Newtonian
        theory, composed of $^{12}$C along sequences of fixed magnetic
        moments including lattice effects in the EoS. The filled dots
        mark the onset of electron capture on $^{12}$C nuclei and the
        threshold electron capture densities for daughter nucleus obtained from
        pycnonuclear reactions are marked with filled squares. The
        corresponding onset for $^{16}$O WDs are shown by open
        symbols. The mass-radius relationship for $^{16}$ O WDs is not
        shown since the difference to $^{12}$C is maximally a few
        percent, hardly visible on the figure.}
    \label{fig:mr_wlat_ecap_12C}
  \end{center}
\end{figure}

\citet{ChamelFantinaDavis} pointed out that the onset of electron captures 
by nuclei (with the emission of a neutrino), 
\begin{equation}\label{eq:e-capture}
^A_ZX+e^- \rightarrow ^A_{Z-1}Y+\nu_e\, ,
\end{equation}
may induce a local instability in magnetic WDs thus limiting their
maximum gravitational mass. In the absence of magnetic fields, the
onset of electron captures occur at mass density
$\rho_\beta \simeq 4.16\times 10^{10}$ g~cm$^{-3}$ (pressure
$P_\beta\simeq 6.99\times 10^{28}$ dyn~cm$^{-2}$) for $^{12}$C and
$\rho_\beta \simeq 2.06\times 10^{10}$ g~cm$^{-3}$
($P_\beta\simeq 2.73\times 10^{28}$ dyn~cm$^{-2}$) for
$^{16}$O~\citep{ChamelPRD92}. In the presence of a strong magnetic
field, the threshold density and pressure are shifted to either higher
or lower values depending on the magnetic field
strength~\citep{ChamelPRD92}, an effect which has been neglected until
recently. \citet{Otoniel} have considered two limiting cases: (i) the
absence of magnetic field, and (ii) the presence of a strongly
quantising magnetic field such that only the lowest Landau level is
filled and $\rho_\beta=\rho_{\rm b}$. In both cases, \citet{Otoniel}
neglected the effects of electron-ion interactions on $\rho_\beta$, as
in the study of \citet{BeraBhatt2015}. In the present work, we take
into account the full dependence of the threshold density $\rho_\beta$
on the magnetic field strength including lattice corrections, as
computed by \citet{ChamelPRD92}. In this way, both the EoS and the
electron capture threshold are calculated fully consistently.

\begin{figure}
  \begin{center}
      \includegraphics[width=.4\textwidth,angle=270]{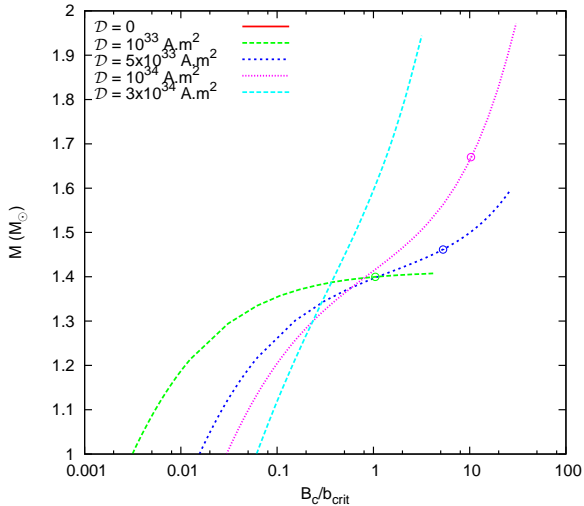}
      \caption{Central magnetic field $B_c$, in units of
        $b_{\rm crit}$~(\ref{eq:Bcrit}), vs WD mass $M$ for different
        fixed magnetic moments $\mathcal{D}$ for Newtonian $^{16}$O
        WDs. The open dots show the onset of electron capture.}
    \label{fig:bcmg_constmm}
  \end{center}
\end{figure}

The resulting mass-radius relations and electron capture thresholds
for magnetic WDs along fixed magnetic moment sequences are shown in
Fig.~\ref{fig:mr_wlat_ecap_12C} considering stars made of either
$^{12}$C or $^{16}$O. The filled and open dots along the sequences
indicate the onset of electron capture instability for $^{12}$C and
$^{16}$O respectively. The mass-radius relations for $^{16}$O differ
only marginally from those obtained for $^{12}$C, and are not
displayed on the figure for better readability. Our calculations
confirm the suggestion of \citet{ChamelFantinaDavis,ChamelPRD90} that
the electron capture instability can be a limiting factor for
determining the maximum WD mass and in turn the maximum magnetic
field. For magnetic moments as large as $10^{34}$~A.m$^2$, electron
capture instability (in the case with lattice) limits the maximum
gravitational mass to 1.86~M$_{\odot}$ for $^{12}$C. For $^{16}$O, the
onset of the instability is reached at even lower values of the
magnetic field, thus leading to a maximum gravitational mass of
1.67~M$_{\odot}$. This can be clearly understood from Fig.~1 of
\citet{ChamelPRD92}, where the onset of electron capture instability
was shown to occur at lower densities for $^{16}$O than for
$^{12}$C. For higher magnetic moments, the star becomes maximally
distorted before the electron capture threshold is reached. In this
case, the maximum gravitational mass could thus be potentially higher
than that shown in Fig.~\ref{fig:mr_wlat_ecap_12C}, see also
\citet{Otoniel}. Fig.~\ref{fig:bcmg_constmm} shows the mass of
$^{16}$O WDs as a function of the core magnetic field strengths for
different fixed magnetic dipole moments. Here we show again the onset
of electron capture instabilities by open dots along the curves. It
is evident from the figure that the higher the magnetic moment is, the
larger is the increase of the mass with the core magnetic field
strength. All in all, no stellar configurations with a poloidal
magnetic field are found with a mass above 2~M$_{\odot}$, if the star
is not allowed to take a toroidal shape.

\begin{table*}
 \centering
 \begin{minipage}{140mm}
   \caption{Maximum masses (in M$_{\odot}$) of strongly magnetised
     $^{16}$O WDs for various magnetic moments (in A.m$^2$), limited
     by different mechanisms: B denotes the onset of torus-like shape
     due to magnetic field, EC stands for electron capture
     instability, pycno is short for pycnonuclear fusion reactions.}
  \begin{tabular}{@{}lrrrrrr@{}}
\hline
  Magnetic moment & & $^{16}$O, without lattice & & & $^{16}$O, with lattice\\
      (A.m$^2$)  & B & B+EC & B+pycno & B & B+EC & B+pycno\\
 \hline
                0 & 1.44 & 1.44 & 1.25 & 1.40 & 1.40 & 1.21\\
        $10^{33}$ & 1.45 & 1.44 & 1.25 & 1.41 & 1.40 & 1.21\\
 $5\times10^{33}$ & 1.60 & 1.50 & 1.25 & 1.59 & 1.46 & 1.22\\
        $10^{34}$ & 2.01 & 1.68 & 1.26 & 1.97 & 1.67 & 1.23\\
 $3\times10^{34}$ & 1.96 & 1.96 & 1.40 & 1.94 & 1.94 & 1.38\\
\hline
\end{tabular}
\label{tab:mmax_O16}
\end{minipage}
\end{table*}

\begin{table*}
 \centering
 \begin{minipage}{140mm}
   \caption{Maximum masses (in M$_{\odot}$) of strongly magnetised
     $^{12}$C WDs for various magnetic moments (in A.m$^2$), limited
     by different mechanisms: B denotes the onset of torus-like shape
     due to magnetic field, EC stands for electron capture
     instability, pycno is short for pycnonuclear fusion reactions.}
  \begin{tabular}{@{}lrrrrrr@{}}
\hline
  Magnetic moment & & $^{12}$C, without lattice & & & $^{12}$C, with lattice\\
      (A.m$^2$)  & B & B+EC & B+pycno & B & B+EC & B+pycno\\
 \hline
                0 & 1.44 & 1.44 & 1.41 & 1.41 & 1.41 & 1.38\\
        $10^{33}$ & 1.45 & 1.44 & 1.41 & 1.41 & 1.41 & 1.38\\
 $5\times10^{33}$ & 1.60 & 1.528 & 1.43 & 1.60 & 1.50 & 1.40\\
        $10^{34}$ & 2.01 & 1.85 & 1.48 & 2.00 & 1.86 & 1.47\\
 $3\times10^{34}$ & 2.03 & 2.03 & 2.03 & 1.96 & 1.96 & 1.96\\
\hline
\end{tabular}
\label{tab:mmax_C12}
\end{minipage}
\end{table*}

\subsection{Pycnonuclear instability}\label{s:pycno}

As first discussed by \citet{ChamelFantinaDavis}, 
pycnonuclear fusion reactions, whereby nuclei transform as 
\begin{equation}
\label{eq:pycno}
^A_Z X + ^A_Z X \to ^{2A}_{2Z} Y \, ,
\end{equation}
could play an important role in limiting the masses of strongly
magnetised WDs. However, the rates of these processes remain highly
uncertain~\citep{YakovlevGasques}. We can estimate the strongest
impact on the maximum mass by considering that pycnonuclear reactions
set in at densities below the threshold density $\rho_{\beta}$ for the
onset of electron capture by the daughter nuclei $^{2A}_{2Z}Y$. From
Table I of \citet{ChamelPRD92}, it is evident that the daughter
nucleus $^{2A}_{2Z}Y$ is generally much more unstable than the parent
nucleus $^A_ZX$ (e.g. the threshold density $\rho_{\beta}$ for
$^{32}$S is $1.69 \times 10^8$ g/cm$^3$ while that of $^{16}$O is
$2.06 \times 10^{10}$ g/cm$^3$, including lattice effects). In
Fig. \ref{fig:mr_wlat_ecap_12C}, we mark the threshold densities
$\rho_{\beta}$ for the daughter nucleus $^{32}$S (from the fusion of
$^{16}$O) and $^{24}$ Mg (from the fusion of $^{12}$C), respectively,
with empty and open squares along the curve. We find that for the
non-magnetic case, the maximum mass decreases from 1.40 M$_{\odot}$ to
1.21 M$_{\odot}$ for $^{16}$O. Excluding lattice interactions
increases both masses by about 3\%. The effect is much less pronounced
for $^{12}$C, where the threshold density is higher. Here the maximum
mass in the non-magnetised case becomes $1.38 M_\odot$. This is in
agreement with the results quoted in \citet{ChamelPRD92}. In the
presence of a strongly quantizing magnetic field, the reduction of the
maximum mass of WDs is even more spectacular: for a magnetic moment of
$10^{34}$ A~m$^2$, the maximum mass decreases from 1.97 M$_{\odot}$ to
1.23 M$_{\odot}$ for $^{16}$O WDs. Because pycnonuclear reactions may
actually occur at densities above the electron capture threshold
density for the daughter nuclei, the values we found for the maximum
mass represent lower bounds.

\section{Gravitational wave emission}\label{s:gw}

It was conjectured by \citet{Heyl} that rotating magnetic WDs could be
important sources of gravitational waves (GW) detectable by the
proposed eLISA mission (Evolved Laser Interferometer Space
Antenna)\footnote{https://elisascience.org}. \citeauthor{Heyl}
estimated the GW emission for a number of observed rotating magnetic
WDs using simple approximations for the reduced quadrupole moment.
Using the analytic prescription of \citet{bonazzola-96}, where the
distortion due to the magnetic field and due to rotation are assumed
to be decoupled, we calculated the gravitational wave amplitudes for
the potential sources listed in Table~1 of \citet{Heyl}. We
numerically estimated the quadrupole moment in a self-consistent way
within our improved microscopic model, assuming a gravitational mass
of $0.6 M_{\odot}$ and an internal magnetic field of $10^{11}$ G. We
obtained a lower value of $0.58 \times 10^{46}$ g cm$^2$ for
quadrupole moment, close to the value $10^{47}$ g cm$^2$ obtained in
\citet{Heyl}. However, although the discussion in \citet{Heyl}
concluded that the amplitude was within the detectable range of the
former LISA interferometer project, our calculations of gravitational
wave amplitudes for the rotating magnetic WDs estimated here are not
likely to be detectable by the current project eLISA, as they are not
within the estimated sensitivity range.


\section{Conclusions}\label{s:conc}
Using the formalism developed by \citet{Chatterjee}, we have studied
the equilibrium structure of WDs endowed with a strong poloidal
magnetic field focusing on the determination of the maximum mass. In
our approach, the coupled equilibrium equations for magnetic and
gravitational fields are solved taking consistently into account
stellar deformations due to rotations and/or anisotropies introduced
by the magnetic field.

For our investigation, we have employed both general relativity and
Newtonian theory for the description of gravity, together with the EoS
of a degenerate electron Fermi gas interacting with a pure ionic
crystal lattice made of $^{12}$C or $^{16}$O. The magnetic-field
dependence of the EoS induced by Landau quantisation of electron
motion has been fully included, as well as the resulting magnetisation
of stellar matter. However, our numerical calculations have
demonstrated that neither of these effects does significantly alter
the structure of magnetised WDs, and they can thus be
neglected. Still, the magnetic field can have a large impact on the
stellar mass and radius. In particular, stellar configurations more
massive than the standard Chandrasekhar limit can be obtained for
magnetic field strengths higher than the critical value
$b_{\rm crit}\sim 4.4\times 10^{13}$~G, as found in previous works. We
have also examined more closely the role of electron-ion interactions,
which have been often neglected in previous studies of strongly
magnetised WDs. Because these interactions are attractive, the maximum
mass is reduced by a few percent. On the other hand, the
stellar radius can be increased by up to $\sim 50\%$ for nonmagnetic
WDs with a mass of 1.4 M$_\odot$, and therefore these interactions
should be included in any realistic model of WDs.

To explore whether strongly magnetised WDs are globally stable, we
have performed calculations of sequences of equilibrium stellar
configurations along fixed values of the magnetic dipole moment. We
have found that the presence of a strong magnetic field can lead to
large deformations of the star. For high enough values of the magnetic
dipole moment, the stellar shape thus becomes toroidal. This extreme
configuration is reached for gravitational masses below $2 M_{\odot}$
and a surface magnetic field of the order of
$10^{13}$-$10^{14}$~G. Note that, with our models of poloidal magnetic
field the core magnetic strength is about one order of magnitude
higher than the surface one. More strongly magnetised stellar
configurations cannot be treated within our numerical framework, as
the star gets a torus-like shape (magnetic pressure becomes larger
than fluid pressure in the centre). We have also investigated the role
of rotation on the stellar structure, and we have found that the
maximum mass of magnetic WDs is increased by a few percent at most,
the effects being the largest in the absence of magnetic fields.

Although torus-shaped strongly magnetised WDs with a mass above
2 M$_{\odot}$ could potentially exist beyond the distorted
configurations computed in the present work, various instabilities can
arise. First of all, it is well-known that general relativity can
limit the maximum mass of nonmagnetic WDs. We have thus computed the
structure of strongly magnetised WDs in full general relativity and
have found that for weakly magnetised WDs, general relativistic
effects reduce the maximum mass, as already discussed by
\citet{ChinPhysB,Coelho,BeraBhatt2015}. On the other hand, general
relativity hardly plays any role in limiting the masses of strongly
magnetised WDs, as anticipated by \citet{KunduBani}. More importantly,
the stability of strongly magnetised WDs is found to be mainly limited
by the onset of electron capture and pycnonuclear reactions in the
stellar core, as argued by \citet{ChamelFantinaDavis}. We have
determined the threshold densities and pressures consistently with the
EoS, including for the first time the effects due to both electron-ion
interactions and the magnetic field. Given the high uncertainties on
the pycnonuclear fusion reaction rates, we have estimated the maximum
possible reduction of the WD mass by assuming that these processes set
in at the same threshold density and pressure as electron captures by
the daughter nuclei. These reactions lead to a drastic decrease in the
maximum WD mass to values even below the Chandrasekhar limit. Our
numerical results about mass limits are summarized in Tables
\ref{tab:mmax_O16} and \ref{tab:mmax_C12}. Additionally, we have
estimated gravitational wave amplitudes emitted from rotating
magnetised WDs, for which the magnetic and rotation axes are
misaligned. We come to the conclusion that, from the currently
observed WDs, none could be detected by the future space-based
gravitational-wave detector eLISA.

Extremely-magnetised WDs can thus in principle reach masses higher
than $2\ M_\odot$ and remain stable with respect to gravitational and
electron capture instabilities; the consequence being that they would
have torus-like shapes. On the other hand, pycnonuclear instablities
could severely limit the existence of such stars, but one must keep in
mind the high uncertainties associated with these reactions. In summary, 
the possibility of super-Chandrasekhar strongly magnetised WDs cannot be 
totally excluded from current theoretical considerations, but important 
issues still need to be addressed before any firm conclusions on 
their existence could be drawn. First, the dynamical stability of these extremely high 
magnetic fields in such cold dense crystallized stellar environment remains to 
be proved. Moreover, the magnetic field strengths expected at the surface of such 
strongly magnetised WDs appear to be four orders of magnitude higher than the 
upper limit of about $10^9$~G set by currently observed magnetic WDs. Finally, 
no realistic and quantitative astrophysical formation scenarios have been so far 
proposed to explain the origin of such strongly magnetised WDs. 


\appendix

\section{General expression for the magnetization}
\label{a:App-EoS}

We report here for completeness and future reference the general
expressions for the magnetization and its first derivative, that are
needed with the EoS to compute the WD structure.
Starting from the definition of $\mathcal{M} \equiv \partial
P/\partial b|_\mu$, Eq.~(\ref{eq:M}), we first observe that constant
chemical potential implies:
\begin{equation}
d\mu = 0 \Longrightarrow d\mu_e = -d\mu_L \ ,
\end{equation}
where we have used Eqs.~(\ref{eq:mu}) and (\ref{eq:mul}).
Writing explicitly the dependence of the pressure on the magnetic
field $b$ and the electron chemical potential $\mu_e$ as $P =
P(b,\mu_e)$, we obtain: 
\begin{equation}
\label{eq:dpdb}
\mathcal{M} \equiv \left.\frac{\partial P}{\partial b}\right|_\mu =
\left.\frac{\partial P}{\partial b}\right|_{\mu_e}  
- \left.\frac{\partial P}{\partial \mu_e}\right|_b
\left.\frac{\partial \mu_L}{\partial b}\right|_{\mu_e}  
\left[ 1 +  \left.\frac{\partial \mu_L}{\partial \mu_e}\right|_b \right]^{-1} \ ,
\end{equation}
with
\begin{equation}
\label{eq:dpdbadd}
\frac{\partial P}{\partial b} = \frac{\partial P_L}{\partial b} +
\frac{\partial P_e}{\partial b} \ , 
\end{equation}
\begin{equation}
\label{eq:dpdmueadd}
\frac{\partial P}{\partial \mu_e} = \frac{\partial P_L}{\partial
  \mu_e} + \frac{\partial P_e}{\partial \mu_e} \ . 
\end{equation}
Using the definition of $b_{\rm crit}$ and $\gamma_e$ from
Eqs.~(\ref{eq:Bcrit}) and (\ref{eq:gammae}), we also note that:
\begin{equation}
  \frac{\partial }{\partial b} = \frac{1}{b_{\rm crit}} \frac{\partial
  }{\partial b_{\star}} \ , 
\end{equation}
\begin{equation}
  \frac{\partial }{\partial \mu_e} = \frac{1}{\xme} \frac{\partial }{\partial \gamma_e} \ .
\end{equation}
We thus find for the first derivative of the pressure (lattice and
electron contribution) with respect to $b$:
\begin{equation}
\label{eq:dpl2db}
\left.\frac{\partial P_L}{\partial b}\right|_{\mu_e} = \frac{4}{9}
\frac{\mathcal{C} e^2}{b_{\rm crit}} 
Z^{2/3} n_e^{1/3} \left.\frac{\partial n_e}{\partial
    b_{\star}}\right|_{\gamma_e} \ , 
\end{equation}
with:
\begin{equation}
\label{eq:dnedb}
\left.\frac{\partial n_e}{\partial b_{\star}}\right|_{\gamma_e} =
\frac{n_e}{b_{\star}} + \frac{2 b_{\star}}{(2 \pi)^2 \lambda_e^3}
\sum_{\nu_L=0}^{\nu_L^{\rm max}} g_\nu \frac{\partial x_e}{\partial
  b_{\star}} \ , 
\end{equation}
\begin{equation}
\label{eq:dxedB}
\frac{\partial  x_e}{\partial  b_{\star}} = -\frac{\nu_L}{x_e} \ ;
\end{equation}

\begin{eqnarray}
\label{eq:dpe2db}
\left.\frac{\partial P_e}{\partial b}\right|_{\mu_e} &=&  \frac{P_e}{b}\nonumber \\
  -
\frac{b_{\star} \xme}{(2\pi)^2 \lambda_e^3 b_{\rm crit}}&\times &
\sum_{\nu_L=0}^{\nu_L^{\rm max}} 2 g_\nu \nu_L \ln \left( \frac{\gamma_e +
    x_e}{\sqrt{1+2\nu_L b_{\star}}} \right) \ . 
\end{eqnarray}
For the first derivative of the pressure (lattice and electron
contribution) with respect to $\mu_e$, we find: 
\begin{equation}
\label{eq:dpl2dmue}
\left.\frac{\partial P_L}{\partial \mu_e}\right|_b = \frac{4}{9}
\frac{\mathcal{C} e^2}{\xme} 
Z^{2/3} n_e^{1/3} \left.\frac{\partial n_e}{\partial
    \gamma_e}\right|_{b_{\star}} \ ,  
\end{equation}
with:
\begin{equation}
\label{eq:dnedgam}
\left.\frac{\partial n_e}{\partial \gamma_e}\right|_{b_{\star}} =
\frac{2 b_{\star}}{(2 \pi)^2 \lambda_e^3} \sum_{\nu_L=0}^{\nu_L^{\rm max}}
g_\nu \frac{\partial  x_e}{\partial  \gamma_e} \ , 
\end{equation}
\begin{equation}
\label{eq:dxedgamma}
\frac{\partial  x_e}{\partial  \gamma_e} = \frac{\gamma_e}{x_e} \ ;
\end{equation}

\begin{equation}
\label{eq:dpe2dmue}
\left.\frac{\partial P_e}{\partial \mu_e}\right|_b =
\frac{b_{\star}}{(2\pi)^2 \lambda_e^3} \sum_{\nu_L=0}^{\nu_L^{\rm max}} 2
g_\nu x_e = n_e \ . 
\end{equation}
The first derivatives of the chemical potential can be written as:
\begin{equation}
\label{eq:dmul2db}
\left.\frac{\partial \mu_L}{\partial b}\right|_{\mu_e} = \frac{4}{9}
\frac{\mathcal{C} e^2}{b_{\rm crit}} Z^{2/3} n_e^{-2/3} \left.\frac{\partial
    n_e}{\partial b_{\star}}\right|_{\gamma_e} \ , 
\end{equation}
using Eq.~(\ref{eq:dnedb}), and

\begin{equation}
\label{eq:dmul2dmue}
\left.\frac{\partial \mu_L}{\partial \mu_e}\right|_b = \frac{4}{9}
\frac{\mathcal{C} e^2}{\xme} 
Z^{2/3} n_e^{-2/3} \left.\frac{\partial
    n_e}{\partial
    \gamma_e}\right|_{b_{\star}} \ , 
\end{equation}
using Eq.~(\ref{eq:dnedgam}).

To calculate the derivative of $M$,
\begin{equation}
\label{eq:dmdmu}
\frac{\partial \mathcal{M}}{\partial \mu} = \frac{\partial ^2
  P}{d\mu \partial b} = \left.\frac{\partial \mathcal{M}}{\partial
    \mu_e}\right|_b \left(\left.\frac{\partial \mu}{\partial
      \mu_e}\right|_b\right)^{-1} \ , 
\end{equation}
we write explicitly the terms appearing in Eq.~(\ref{eq:dmdmu}) as:
\begin{eqnarray}
\left.\frac{\partial \mathcal{M}}{\partial \mu_e}\right|_b &=&
\frac{1}{\xme} \left\{ \frac{\partial }{\partial \gamma_e}
  \left.\frac{\partial P}{\partial b}\right|_{\mu_e} \right. \nonumber
\\ 
   &-& \left. \frac{\partial }{\partial \gamma_e}
     \left(\left.\frac{\partial P}{\partial
           \mu_e}\right|_{b_{\star}}\right) \left.\frac{\partial
         \mu_L}{\partial b}\right|_{\mu_e} \left( 1 +
       \left.\frac{\partial \mu_L}{\partial \mu_e}\right|_b
     \right)^{-1} \right. \nonumber \\ 
   &-& \left. \left.\frac{\partial P}{\partial
         \mu_e}\right|_{b_{\star}} \frac{\partial }{\partial \gamma_e}
     \left(\left.\frac{\partial \mu_L}{\partial
           b}\right|_{\mu_e}\right) \left( 1 + \left.\frac{\partial
           \mu_L}{\partial \mu_e}\right|_b \right)^{-1}
   \right. \nonumber \\ 
   &+& \left. \left.\frac{\partial P}{\partial
         \mu_e}\right|_{b_{\star}} \left.\frac{\partial
         \mu_L}{\partial b}\right|_{\mu_e} \left( 1 +
       \left.\frac{\partial \mu_L}{\partial \mu_e}\right|_b
     \right)^{-2} \right. \nonumber \\ 
   &\times & \left. \frac{\partial }{\partial \gamma_e} \left(
       \left.\frac{\partial \mu_L}{\partial
           \mu_e}\right|_{b_{\star}}\right) \right\} \ , 
\end{eqnarray}
and
\begin{equation}
\left.\frac{\partial \mu}{\partial \mu_e}\right|_b = y_e \left( 1+
  \left.\frac{\partial \mu_L}{\partial \mu_e}\right|_b \right) \ , 
\end{equation}
where each contribution of the pressure derivative can be calculated
according to Eqs.~(\ref{eq:dpdbadd})-(\ref{eq:dpdmueadd}). We thus
obtain, for the second derivative of the pressure (electron and
lattice contribution):
\begin{eqnarray}
\label{eq:d2pl2dbdgam}
\frac{\partial }{\partial \gamma_e} \left(\left.\frac{\partial
      P_L}{\partial b}\right|_{\mu_e}\right)_{b_{\star}}\mkern-18mu &=&
\frac{4}{9} \frac{\mathcal{C} e^2}{b_{\rm crit}} Z^{2/3} \left[
  \frac{1}{3} n_e^{-2/3} \left.\frac{\partial n_e}{\partial
      \gamma_e}\right|_{b_{\star}} \left.\frac{\partial
      n_e}{db_{\star}}\right|_{\gamma_e} \right. \nonumber \\ 
&+& \left. n_e^{1/3} \frac{\partial }{\partial \gamma_e}
  \left.\left(\frac{\partial n_e}{\partial
        b_{\star}}\right)\right|_{b_{\star}} \right] \ , 
\end{eqnarray}
using Eqs.~(\ref{eq:dnedb}) and (\ref{eq:dnedgam}), and
\begin{eqnarray}
\label{eq:d2nedbdgam}
\frac{\partial }{\partial \gamma_e} \left(\left.\frac{\partial
      n_e}{\partial b_{\star}}\right|_{\mu_e}\right)_{b_{\star}} &=&
\frac{2}{(2 \pi)^2 \lambda_e^3} \sum_{\nu_L=0}^{\nu_L^{\rm max}} g_\nu
\frac{\partial  x_e}{\partial  \gamma_e} \nonumber \\ 
 &+& \frac{2 b_{\star}}{(2 \pi)^2 \lambda_e^3} \sum_{\nu_L=0}^{\nu_L^{\rm
     max}} g_\nu \frac{\partial }{\partial \gamma_e} \left(
   \frac{\partial  x_e}{\partial  b_{\star}} \right)\ , 
\end{eqnarray}
with $\partial  x_e/\partial  \gamma_e$ from Eq.~(\ref{eq:dxedgamma}) and
\begin{equation}
\label{eq:d2xedgammadB}
\frac{\partial }{\partial \gamma_e} \left( \frac{\partial
    x_e}{\partial  b_{\star}} \right) = \frac{\nu_L \ \gamma_e}{x_e^3} \
; 
\end{equation}

\begin{equation}
\label{eq:d2pe2dbdgam}
\frac{\partial }{\partial \gamma_e} \left( \left.\frac{\partial
      P_e}{\partial b}\right|_{\mu_e}\right)_{b_{\star}} = \frac{n_e
  \xme}{ b} 
- \frac{b_{\star} \xme}{(2 \pi)^2 \lambda_e^3 b_{\rm crit}}
\sum_{\nu_L=0}^{\nu_L^{\rm max}} 2 \frac{g_\nu \nu_L}{x_e} \ ; 
\end{equation}

\begin{eqnarray}
\label{eq:d2pldmuedgam}
\frac{\partial }{\partial \gamma_e} \left(\left.\frac{\partial
      P_L}{\partial \mu_e}\right|_{b_{\star}}\right)_{b_{\star}} &=&
\frac{4}{9} \frac{\mathcal{C} e^2}{\xme} Z^{2/3} \left[ \frac{1}{3}
  n_e^{-2/3} \left(\left.\frac{\partial n_e}{\partial
        \gamma_e}\right|_{b_{\star}}\right)^2 \right. \nonumber \\ 
&+& \left. n_e^{1/3} \frac{\partial }{\partial \gamma_e}
  \left(\left.\frac{\partial n_e}{\partial
        \gamma_e}\right|_{b_{\star}}\right) \right] \ , 
\end{eqnarray}
using Eq.~(\ref{eq:dnedgam}) and:
\begin{eqnarray}
\label{eq:d2nedgamdgam}
\frac{\partial }{\partial \gamma_e} \left(\left.\frac{\partial
      n_e}{\partial \gamma_e}\right|_{b_{\star}}\right)_{b_{\star}} =
\frac{2 b_{\star}}{(2\pi)^2 \lambda_e^3} \sum_{\nu_L=0}^{\nu_L^{\rm max}}
g_\nu \frac{\partial }{\partial  \gamma_e} \left( \frac{\partial
    x_e}{\partial  \gamma_e} \right) \ , 
\end{eqnarray}
with:
\begin{equation}
\frac{\partial }{\partial  \gamma_e} \left( \frac{\partial
    x_e}{\partial  \gamma_e} \right) = -\frac{1+2\nu_L b_{\star}}{x_e^3}
\ ; 
\end{equation}

\begin{equation}
\label{eq:d2pedmuedgam}
\frac{\partial }{\partial \gamma_e} \left(\left.\frac{\partial
      P_e}{\partial \mu_e}\right|_{b_{\star}}\right)_{b_{\star}} =
\frac{2 b_{\star}}{(2\pi)^2 \lambda_e^3} 
   \sum_{\nu_L=0}^{\nu_L^{\rm max}} \frac{g_\nu \ \gamma_e}{x_e} \ .
\end{equation}
For the second derivatives of the chemical potential, we have:
\begin{eqnarray}
\label{eq:d2muldbdgam}
\frac{\partial }{\partial \gamma_e} \left(\left.\frac{\partial
      \mu_L}{\partial b}\right|_{\mu_e}\right)_{b_{\star}} &=&
\frac{4}{9} \frac{\mathcal{C} e^2}{b_{\rm crit}} Z^{2/3} \nonumber \\ 
&\times & \left[ -\frac{2}{3} n_e^{-5/3} \left(\left.\frac{\partial
        n_e}{\partial \gamma_e}\right|_{b_{\star}}\right)
  \left(\left.\frac{\partial n_e}{\partial
        b_{\star}}\right|_{\mu_e}\right) \right. \nonumber \\ 
&+& \left. n_e^{-2/3} \frac{\partial }{\partial \gamma_e}
  \left(\left.\frac{\partial n_e}{\partial
        b_{\star}}\right|_{\mu_e}\right)_{b_{\star}} \right] \ , 
\end{eqnarray}
where we have used Eqs.~(\ref{eq:dnedgam}) and (\ref{eq:d2nedbdgam}), and

\begin{eqnarray}
\label{eq:d2muldmuedgam}
\frac{\partial }{\partial \gamma_e} \left(\left.\left(\frac{\partial
        \mu_L}{\partial
        \mu_e}\right)\right|_{b_{\star}}\right)_{b_{\star}} &=&  
 \frac{4}{9} \frac{\mathcal{C} e^2}{\xme} Z^{2/3} \nonumber \\
 &\times & \left[ -\frac{2}{3} n_e^{-5/3} \left(\left.\frac{\partial
         n_e}{\partial \gamma_e}\right|_{b_{\star}}\right)^2
 \right. \nonumber \\ 
 &+&  \left. n_e^{-2/3} \frac{\partial }{\partial \gamma_e}
   \left(\left.\frac{\partial n_e}{\partial
         \gamma_e}\right|_{b_{\star}}\right)_{b_{\star}} \right] \ , 
\end{eqnarray}
where we made use of Eqs.~(\ref{eq:dnedgam}) and
(\ref{eq:d2nedgamdgam}). 

\section{Derivation of the first integral in general relativity}\label{a:App}
The constraint for the conservation of energy and momentum in equilibrium
can is expressed as :
\begin{equation}
\nabla_\mu T^{\mu\nu} = 0.
\end{equation}

Inserting the expression for the energy momentum tensor as in Eq.(\ref{eq:tmunu})
into the above equation yields \citep{Chatterjee}
\begin{eqnarray}
  (\mathcal{E} + P) \left ( \frac{1}{\mathcal{E} + P} \frac{\partial
      P}{\partial x^i} + \frac{\partial \nu}{\partial x^i} -
    \frac{\partial \ln \Gamma}{\partial x^i}\right) && \nonumber \\  
  -  F_{i \rho} j^\rho_{\ \mathit{free}} 
  - \frac{\chi}{2\mu_0}  F_{\mu\nu}
  \nabla_i F^{\mu\nu} &=& 0~.
\label{eq:first-integral}
\end{eqnarray}
where one can identify the first part as the perfect-fluid contribution to
the energy-momentum tensor and the Lorentz force
terms arising from free currents and the magnetization.
Defining the enthalpy as a function of both baryon density and
magnetic field norm $b = \sqrt{b^\mu b_\mu}$ in the comoving frame:
\begin{equation}
 h = h(n, b) = \frac{\mathcal{E} + P}{m_B n c^2}~, \label{e:def_h}
\end{equation} 
where $m_B$ is the mean mass of a baryon. One can rewrite the first
term in the parentheses of Eq. (\ref{eq:first-integral}) in terms of
the enthalpy as
\begin{equation}
\frac{\partial P}{\partial x^i} = (\mathcal{E}+P)\frac{\partial \ln
  h}{\partial x^i} + \mathcal{M} \frac{\partial b}{\partial x^i}\, . 
\label{eq:cancel1}
\end{equation}
The magnetic field in the comoving frame $b^\mu$ is defined by 
\begin{equation}
F_{\mu\nu} = \epsilon_{\alpha\beta\mu\nu} u^{\beta} b^{\alpha}
\label{eq:deffmunu} 
\end{equation}
and
\begin{equation}
{\mathcal M}_{\mu\nu} = \epsilon_{\alpha\beta\mu\nu} u^{\beta} \mathcal{M}^{\alpha}.
\label{eq:defmmunu}
\end{equation}

Further, the last term in  Eq. (\ref{eq:first-integral}) can be
written in terms of the magnetic field $b^\mu$ in 
the comoving frame as :
\begin{eqnarray}
  \frac{\chi}{2\mu_0} F_{\mu\nu} \nabla_i F^{\mu\nu} &=&
  \frac{\chi}{\mu_0}b_\mu \nabla_i b^\mu - b_\mu b^\mu u_\nu \nabla_i
  u^\nu \nonumber \\
  &=& b \nabla_i b = \mathcal{M} \df{b}{x^i}.
\label{eq:cancel2}
\end{eqnarray} 
Thus, inserting Eqs. (\ref{eq:cancel1}) and (\ref{eq:cancel2}) into
Eq. (\ref{eq:first-integral}), one gets an expression without explicit
dependence on magnetization
\begin{equation}
\frac{\partial \ln h}{\partial x^i} + \frac{\partial \nu}{\partial x^i}
- \frac{\partial \ln \Gamma}{\partial x^i}  -  \frac{F_{i \rho}
  j^\rho_{\ \mathit{free}}}{\mathcal{E} + P}  = 0,
\end{equation}
which has the same form as in absence of magnetization \citep{BGSM,
  Bocquet}. Finally, as in \citet{Chatterjee} we introduce the
\textsl{current function}, whose primitive is noted $\Phi(r,\theta)$
and defined by Eq.~\eqref{e:def_Phi}. This enables us to get a first
integral of motion as given in Eq.~\eqref{eq:1st_integral}.

\section*{Acknowledgements}
We warmly thank M.~Mouchet and E.~Gourgoulhon for fruitful
discussions. D.C. would like to thank the Observatoire de Paris in
Meudon, particularly the computing department, for their constant
support. This work has been funded by the SN2NS project
ANR-10-BLAN-0503, the ``Gravitation et physique fondamentale'' action
of the Observatoire de Paris, Fonds de la Recherche Scientifique -
FNRS (Belgium) and the COST action MP1304 ``NewCompStar''. 

\label{lastpage}

\end{document}